\documentstyle[preprint,aps]{revtex}
\tighten
\onecolumn
\begin{document}
\title{Greenberger-Horne-Zeilinger paradoxes for multiport
beam splitters}
\author{
Marek \.Zukowski and Dagomir Kaszlikowski}
\address{Instytut Fizyki Teoretycznej i Astrofizyki\\
Uniwersytet Gda\'nski, PL-80-952 Gda\'nsk, Poland}
\maketitle
\begin{abstract}
In a gedankenexperiment $N$ particles in a generalised GHZ-type beam
entangled state (each particle can be in one of $M$ beams) are fed
into $N$ symmetric $2M$-port beam splitters (spatially separated).
Correlation functions for such a process (using the Bell numbers value
assignment approach) reveal a remarkable symmetry. For $N=M+1\geq 4$ a
series of GHZ paradoxes are shown.

\end{abstract}

The ideas of Einstein, Podolsky and Rosen \cite{EPR} are based on the
observation that for some systems quantum mechanics predicts perfect
correlations of their properties. Greenberger, Horne and Zeilinger
(GHZ) have shown, that such correlations, in case of three or more
entangled particles, lead to a strikingly more direct refutation of
the argument of EPR on the possibility of introducing elements of
reality to complete quantum mechanics, than considerations involving
only pairs of particles \cite{GHZ}, \cite{GHSZ}. Simply, in the
entangled GHZ state, such correlations cannot be consistently used to
infer at a distance hidden properties of the particles. In
contradistinction to the original two particle Bell theorem, the idea
of EPR, to turn the exact predictions of quantum mechanics against the
claim of its completeness, breaks down already at the stage of
defining the elements of reality.

So far the analysis of GHZ correlations was constrained to dichotomic
observables (for each of the particles). Within this constraint, it
was shown, with the use of various $N$-particle Bell inequalities
that, the violation of the premises of local realism grows with the
number of particles \cite{MERMIN}. I.e, the increasing number of
particles, in this case, does not bring us closer to the classical
realm, as it is often supposed, but rather, makes the discrepancies
between the quantum and the classical more profound.

In this paper we would like to examine whether GHZ-type paradoxes exist
also in the case of correlations expected in gedankenexperiments
involving multiport beam splitters \cite{earlier}, i.e. for a specific
case of nondichotomic observables (which have properties distinctive
to the dichotomic ones \cite{non}). To this end, we shall study a
GHZ-Bell type experiment in which one has a source emitting
$N$-particles in a specific  entangled state of the property, that the
particles propagate towards one of $N$ spatially separated
non conventional measuring devices operated by independent observers.
Each of the devices consists of a symmetric multiport beam splitter
\cite{ZZH} (with $M$ input and $M$ exit ports), $M$ phase shifters
operated by the observers (one in front of each input), and $M$
detectors (one behind each exit port).

Symmetric $2M$-port beam splitter is defined as an $M$-input and
$M$-output interferometric device which has the property that a beam
of light entering via single port is evenly split between all output
ports. I.e., the unitary matrix defining such a device has the
property that
the modulus of all its elements equals ${1\over \sqrt M}$. An extended
introduction to the physics and theory of such devices is given in
\cite{ZZH}, and therefore the reader not acknowledged with those
concepts is kindly asked to consult this reference. Multiport
beam splitters were introduced into the literature on the EPR paradox
in \cite{earlier} in order to extend two particle Bell-phenomena to
observables described as operators in Hilbert spaces of dimension
higher than two. In contradistinction to the higher than 1/2 spin
generalisations of the Bell-phenomena \cite{MS}, this type of
experimental devices generalise the ideas of beam-entanglement
\cite{HZ}. Symmetric multiport beam splitters are performing unitary
transformations between "mutually unbiased" bases in the Hilbert space
\cite{Ivanovic}. They were tested in several recent experiments
\cite{PHD} \cite{MMWZZ}, and also various aspects of such devices were
analysed theoretically \cite{RECK} \cite{Stenholm}.

We shall use here only multiport beam splitters which have the property
that the elements of the unitary transformation which describes their
action are given by
\begin{equation}
U_{m,m'}^{M}={1\over \sqrt M}\gamma_{M}^{(m-1)(m'-1)},
\end{equation}
where $\gamma_{M}=\exp(i{2\pi\over M})$ and the indices $m$, $m'$
denote the input and exit ports. Such devices were called in
\cite{ZZH} the Bell multiports.

We assume that the initial $N$ particle state that feeds $N$ spatially
separated multiports, each of which has $M$ inputs and $M$ outputs,
has the following form:

\begin{eqnarray}
\label {eq1}
&&|\psi(N)\rangle={1\over \sqrt
M}\sum_{m=1}^{M}\prod_{l=1}^{N}|m\rangle_{l},
\label{2}
\end{eqnarray}
where $|m\rangle_{l}$ describes the $l$-th particle being in the
$m$-th beam, which leads to the $m$-th input of the $l$-th multiport.
Please note, that only one particle enters each multiport. However,
However, each of the particles itself is in a mixed state (with equal
weights), which gives it equal probability to enter the local
multiport via any of the input ports.

The state (\ref{2}) seems to be the most straightforward
generalisation of the GHZ states to the new type of observables. In
the original GHZ states the number of their components (i.e., two) is
equal to the dimension of the Hilbert space describing the relevant
(dichotomic) degrees of freedom of each of the particles. This
property is shared with the EPR-type states proposed in \cite{ZZH} for
a two-multiport Bell-type experiment - in this case the number of
components equals the number of input ports of each of the multiport
beam splitters. We shall not discuss here the possible methods to
generate such states. However, we briefly mention that the recently
tested entanglement swapping \cite{ZZHE} technique could be used for
this purpose.

As it was mentioned earlier, in front of every input of each multiport
beam splitter one has a tunable phase shifter. The initial state is
transformed by the phase shifters into

\begin{eqnarray}
\label {eq2}
&&|\psi(N)'\rangle={1\over \sqrt M}\sum_{m=1}^{M}\prod_{l=1}^{N}
\exp(i\phi_{l}^{m})|m\rangle_{l},
\end{eqnarray}
where $\phi_{l}^{m}$ stands for the setting of the phase shifter in
front of the $m$-th port of the $l$-th multiport.

The quantum prediction for  probability to register the first photon
in the output $k_{1}$ of an $2M$ - port device, the second one in the
output $k_{2}$ of the second such device ,..., and the $N$-th one in
the output $k_{N}$ of the $N$-th device is given by:
\begin{eqnarray}
 &P_{QM}(k_{1},\cdots,k_{N}|\vec{\phi_{1}},\cdots,\vec{\phi_{N}})=&\nonumber\\
 &({1\over M})^{N+1}|\sum_{m=1}^{M}\exp(i\sum_{l=1}^{N}\phi_{l}^{m})
 \prod_{n=1}^{N}\gamma_{M}^{(m-1)(k_{n}-1)}|^{2}=&\nonumber\\
 &=({1\over M})^{N+1}\left[M+2\sum_{m>m'}^{M}
 \cos\left(\sum_{l=1}^{N}\Delta\Phi_{l,k_{l}}^{m,m'}
 \right)\right],&
\label{5}
\end{eqnarray}
where
$\Delta\Phi_{l,k_{l}}^{m,m'}=\phi_{l}^{m}-\phi_{l}^{m'}+{2\pi\over
M}(k_{l}-1)(m-m')$, and the indices $m$, $m'$ in the last expression
are understood as expressed modulo M. The shorthand symbol
$\vec{\phi}_{k}$ stands for the full set of phase settings in front of
the $k$-th multiport, i.e.
$\phi_{k}^{1},\phi_{k}^{2},\cdots,\phi_{k}^{M}$.

Let us employ a specific value assignment method (called Bell
number assignment; for a detailed
explanation see again \cite{ZZH}), which ascribes to the detection event
behind the $m$
- th output of a multiport the value  $\gamma_{M}^{m-1}$, where
$\gamma_{M}=\exp(i{2\pi\over M})$.
With such a value assignment to the detection events, the Bell-type
correlation function, which is the average of the product of the
expected results, is defined as

\begin{eqnarray}
&E(\vec{\phi_{1}},\cdots,\vec{\phi_{N}})=&\nonumber\\
&=\sum_{k_{1},\cdots,k_{N}=1}^{M}
\prod_{l=1}^{N}\gamma_{M}^{k_{l}-1}P(k_{1},\cdots,k_{N}|\vec{\phi_{1}},\cdots,\vec{\phi_{N}}).&
\label{eq4}
\end{eqnarray}

The easiest way to compute the correlation function for the quantum
prediction employs the mid formula of (\ref{5}):
\begin{eqnarray}
&E_{QM}(\vec{\phi_{1}},\cdots,\vec{\phi_{N}})=&\nonumber\\ &=({1\over
M})^{N+1}\sum_{k_{1},\cdots,k_{N}=1}^{M}\sum_{m,m'=1}^{M}
\exp(i\sum_{n=1}^{N}(\phi_{n}^{m}-\phi_{n}^{m'}))&\nonumber\\
&\times \prod_{l=1}^{N}\gamma_{M}
^{(k_{l}-1)(m-m'+1)}=&\nonumber\\
&=({1\over M})^{N+1}\sum_{m,m'=1}^{M}
\exp(i\sum_{n=1}^{N}(\phi_{n}^{m}-\phi_{n}^{m'}))&\nonumber\\
&\times\prod_{l=1}^{N}
\sum_{k_{l}=1}^{M}\gamma_{M}
^{(k_{l}-1)(m-m'+1)}.&
\label{corr1}
\end{eqnarray}
Now, one notices that
$\sum_{k_{l}=1}^{M}\gamma_{M}^{(k_{l}-1)(m-m'+1)}$
differs from zero
(and equals to M)
only if $m-m'+1=0$, modulo M. Therefore we can finally write:

\begin{eqnarray}
&E_{QM}(\vec{\phi_{1}},\cdots,\vec{\phi_{N}})=\nonumber&\\ &={1\over
M}\sum_{m=1}^{M}\exp(i\sum_{l=1}^{N}\phi^{m,m+1}_{l}),&
\label{corr2}
\end{eqnarray}
where $\phi^{m,m+1}_{l}=\phi^{m}_{l}-\phi^{m+1}_{l}$ and the above
sum is
understood modulo M, which means that
$\phi^{M+1}_{l}=\phi^{1}_{l}$.

One can notice here a striking simplicity and symmetry of this quantum
correlation function (\ref{corr2}). It is valid for all possible
values of N (number of particles) and for all possible values of
$M\geq2$ (number of ports). For $M=2$, it reduces itself to the usual
two particle, and for $M=2$, $N\geq2$ the standard GHZ type
multiparticle correlation function for beam-entanglement experiments,
namely $\cos(\sum_{l=1}^{N}\phi_{l}^{1,2})$ \cite{MERMIN}. The Bell -
EPR phenomena discussed in
 \cite{ZZH} are described by (\ref{corr2}) for $N=2, M\geq 3$.

 Even for $M=2$, $N=1$ the function (\ref{corr2}) has an interpretation
 for the following process. Assume that
a traditional four-port 50-50 beam splitter, is fed a single photon
input in  a state in which is an equal superposition of being in each
of the two input ports. The value of (\ref{corr2}) is the average of
expected photo counts behind the exit ports (provided the click at one
of the detectors is described as $+1$ and at the other one as $-1$),
and of course it depends of the relative phase shifts in front of the
beam splitter. In other words, this situation describes a Mach-Zehnder
interferometer with a single photon input at a chosen input port. For
$M=3$, $N=1$ the same interpretation applies to the case of a
generalised three input, three output Mach-Zehnder interferometer
described in \cite{HARALD}, provided one one ascribes to firings of
the three detectors respectively
$\gamma_{3}=\alpha\equiv\exp(i{2\pi\over 3})$, $\alpha^{2}$ and
$\alpha^{3}$.

The described set of gedankenexperiments is rich in EPR-GHZ
correlations (for $N\geq2$). To reveal the above, let us first analyse
the conditions (i.e. settings) for such correlations. As the
correlation function (\ref{corr2}) is an average of complex numbers of
unit modulus, one has
$|E_{QM}(\vec{\phi_{1}},\cdots,\vec{\phi_{M})}|\leq 1$. The equality
signals a perfect EPR-GHZ correlation. It is easy to notice that this
may happen only if
$$\exp(i\sum_{l=1}^{N}\phi_{l}^{1,2})=
\exp(i\sum_{l=1}^{N}\phi_{l}^{2,3})=\cdots=
\exp(i\sum_{l=1}^{N}\phi_{l}^{M,1})=\gamma_{M}^{k},$$ where k is an arbitrary
natural number. Under this condition
$E(\phi_{1},\cdots,\phi_{N})=\gamma_{M}^{k}$. This means that only
those sets of $N$ spatially separated detectors may fire, which are
ascribed such Bell numbers which have the property that their product
is $\gamma_{M}^{k}$. Knowing, which detectors fired in a  the set of
$N-1$ observation stations, one can predict with certainty which
detector would fire at the sole observation station not in the set.

We shall now present the simplest GHZ-type paradox for such systems.
We take $M=3$ and $N=4$. That is we consider now, the experimental
situation in which one has the source producing the ensemble of four
three-state particles described by the state $|\psi(4)\rangle$
(compare, (\ref{2})) that feeds four three-port beam splitters (i.e.,
tritters \cite{ZZH}). In this case the quantum correlation function
has the form:

\begin{eqnarray}
&E_{QM}(\vec{\phi_{1}},\vec{\phi_{2}},\vec{\phi_{3}},\vec{\phi_{4}})
=&\nonumber\\
&={1\over 3}\sum_{k=1}^{3}\exp(i\sum_{l=1}^{4}(\phi_{l}^{k}-\phi_{l}^{k+1})).&
\end{eqnarray}

The (deterministic) local hidden variable (L.H.V.) correlation
function for this type of experiment must have the following structure
\cite{BELL}:

\begin{eqnarray}
&E_{HV}(\vec{\phi_{1}},\vec{\phi_{2}},\vec{\phi_{3}},\vec{\phi_{4}})
=&\nonumber\\
&=\int_{\Lambda}\prod_{k=1}^{4}I_{k}(\vec{\phi_{k}},\lambda)\rho(\lambda)d\lambda.&
\end{eqnarray}
The hidden variable function $I_{k}(\vec{\phi_{k}},\lambda)$, which
determines the firing of the detectors behind the k-th multiport,
depends only upon the local set of phases, and takes one of the three
possible values $\alpha$, $\alpha^{2}$, $\alpha^{3}=1$ (these values
indicate which of the detectors is to fire), and $\rho(\lambda)$ is
the distribution of hidden variables.


Consider four gedanken experiments. In the first one our observers,
each of whom operates one of the spatial separated devices,  choose
the following phases in front of their three-port beam splitters:

\begin{eqnarray}
&\vec{\phi_{1}}\equiv(\phi_{1}^{1},\phi_{1}^{2},\phi_{1}^{3})=(0,{2\pi\over 9},{4\pi\over 9})=\vec{\phi_{2}}
=\vec{\phi_{3}}=\vec{\phi}&
\nonumber\\
&\vec{\phi_{4}}\equiv(\phi_{4}^{1},\phi_{4}^{2},\phi_{4}^{3})=(0,0,0)=\vec{\phi'}.&
\end{eqnarray}
In the second experiment, the third observer  sets
$\vec{\phi_{3}}=\vec{\phi'}$ whereas the other ones set $\vec{\phi}$.
We repeat this swapping of the settings procedure in  the next two
experiments until the first observer sets $\vec{\phi'}$ and the other
set $\vec{\phi}$. Quantum mechanics predicts that in all four such
experiments 
the correlation function
is equal to $\alpha^{2}$ (i.e. we have perfect GHZ correlations).
Namely we have
\begin{eqnarray}
&E_{QM}(\vec{\phi},\vec{\phi},\vec{\phi},\vec{\phi}')
=E_{QM}(\vec{\phi},\vec{\phi},\vec{\phi}',\vec{\phi})&\nonumber\\
&=E_{QM}(\vec{\phi},\vec{\phi}',\vec{\phi},\vec{\phi})=
E_{QM}(\vec{\phi}',\vec{\phi},\vec{\phi},\vec{\phi})=\alpha^2.&
\label{q}
\end{eqnarray}

However, this immediately implies that for any L.H.V. theory that aims
at describing these phenomena one must have for every $\lambda$

\begin{eqnarray}
&I_{k}(\vec{\phi'},\lambda)\prod_{{l=1},{l\neq
k}}^4I_{l}(\vec{\phi},\lambda)
=\alpha^{2},&
\label{hidden1}
\end{eqnarray}
and this must hold for all $k=1,2,3,4$. But, since
$I_{l}(\vec{\phi},\lambda)^3=\alpha^{3k}=1$ (where, $k$ represents a
certain integer), then after multiplying these four equations side by
side, one has for every $\lambda$

\begin{eqnarray}
&\prod_{l=1}^{4}I_{l}(\vec{\phi'},\lambda)=\alpha^{2}.&
\end{eqnarray}
Therefore, if the local hidden variable theory is to agree with the
earlier mentioned quantum predictions (\ref{q}), then one must have

\begin{eqnarray}
&E(\vec{\phi'},\vec{\phi'},\vec{\phi'},\vec{\phi'})=\alpha^{2}=\alpha^{*}.&
\label{contradiction}
\end{eqnarray}
However, the quantum prediction is
$E_{QM}(\vec{\phi'},\vec{\phi'},\vec{\phi'},\vec{\phi'})=1$. Thus we
have a GHZ-type contradiction that $1=\alpha^{*}$. I.e., hidden
variables predict a different type perfect EPR-GHZ correlation.


We will extend the reasoning to the case when one has $N$ particles
(described by the state of the form (\ref{2})) beamed into multiport
beam splitters with $M=N-1$. The quantum prediction for the Bell
correlation function is given by     (\ref{corr2}), with the
appropriate value of $M$.

 The L.H.V correlation function must have
the following structure:

\begin{eqnarray}
&\int\prod_{k=1}^{N}I_{k}(\vec{\psi_{k}},\lambda)\rho(\lambda)d\lambda,
\end{eqnarray}
where $k$ now extends from $1$ to $N$ and $\vec{\psi_{k}}$ stands for the
full set of settings in front of the $k$-th multiport, i.e.
$\psi_{k}^{1},
\psi_{k}^{2},\cdots,\psi_{k}^{N-1}$, and $I_{k}(\vec{\psi_{k}},\lambda)$
is a  hidden variable function depending on the local phase settings,
which has the property that its value, which can be any integer power
of $\gamma_{N-1}$, indicated which local detector is to fire.

Now, as it was in the previous case, we must choose appropriate phases
for each of observers that will be taken in the first experiment. The
appropriate choice is the following one:

\begin{eqnarray}
&\vec{\psi}_{1}=\cdots=\vec{\psi}_{N-1}=
(0,\delta,2\delta,\cdots,(N-2)\delta)=\vec{\psi}&\nonumber\\
&\vec{\psi}_{N}=(0,\cdots,0)=\vec{\psi'},&
\end{eqnarray}
where $\delta={2\pi\over (N-1)^2}$. In the next $N-1$ experiments one
applies previously described swapping of the settings procedure until
the first observer sets $\vec{\psi'}$ and the other ones set
$\vec{\psi}$. For such choice of phases the quantum correlation
function for every of the $N$ experiment, is equal to
$\gamma_{N-1}^{N-2}=\gamma_{N-1}^*$.
(i.e. we have perfect GHZ correlations of the same type for each of
the experiments). But this implies that, for any L.H.V. theory that
aims at describing these phenomena one must have, for every $\lambda$,

\begin{eqnarray}
&I_{k}(\vec{\psi'},\lambda)\prod_{l\neq k}I_{l}(\vec{\psi},\lambda)
=\gamma_{N-1}^{*},&
\label{hidden2}
\end{eqnarray}
and that this must hold for all $k=1,\cdots,N$. However, after
multiplying these $N$ equations one has:

\begin{eqnarray}
&\prod_{l=1}^{N}I_{l}(\vec{\psi'},\lambda)=\gamma_{N-1}^{*},&
\end{eqnarray}
where we have used the property of the Bell numbers generated by
$\gamma_{N-1}$, that each of them to the $N-1$-th power gives 1, and
therefore that $I_{k}(\vec{\psi},\lambda)^{N}=1$. Thus, the local
hidden variable implies that
$$E_{QM}(\vec{\psi'},\cdots,
\vec{\psi'})=\gamma_{N-1}^{*}.$$
However, the quantum prediction is 1. Thus we have the GHZ
contradiction that $1=\gamma_{N-1}^{*}$. I.e. hidden variables predict
a different type perfect EPR - GHZ correlations. In other words the
EPR idea of elements of reality makes no sense for the discussed
experiments, and this hold for an arbitrary number of particles $N$,
and for suitably related ($N-1$), but in principle arbitrarily high
number of input and exit ports of  symmetric multiport beam splitters.

In conclusion we state that the multiport beam splitters, and the idea
of value assignment based Bell numbers, lead to a strikingly
straightforward generalisation of the GHZ paradox for non-dichotomic
observables. These properties may possibly find an application in
future quantum information and communication schemes (especially as
GHZ states are now observable in the lab \cite{Innsbruck}).

MZ was supported by the University of Gdansk Grant No
BW/5400-5-0202-8. DK was supported by the KBN Grant 2 P03B 096 15.


\begin{references}

\bibitem{EPR}
  A. Einstein, B. Podolsky, and N. Rosen, { Phys. Rev.} {\bf 47}
 (1935) 777.
\bibitem{GHZ}
D.M. Greenberger, M.A. Horne and A. Zeilinger, in {\it Bell's theorem
and the Conception of the Universe}, edited by M. Kafatos (Kluwer
Academic, Dordrecht, 1989).
\bibitem{GHSZ}
 D.M. Greenberger, M. Horne, A. Shimony and  A.  Zeilinger,
1990,  Am.  J.
Phys., {\bf 58}, 1131.
\bibitem{MERMIN}
N.D. Mermin, Phys. Rev. Lett. {\bf 65}, 1838 (1990); S.M. Roy and V.
Singh, Phys. Rev. Lett. {\bf 67}, 2761 (1991); M. Ardehali, Phys. Rev.
A {\bf 46}, 5375 (1992); M. \.Zukowski, Phys. Lett. A, {\bf 177}, 290
(1993); M. \.Zukowski and D. Kaszlikowski, Phys. Rev. A, {\bf 56},
R1682 (1997).
\bibitem{earlier} D.N. Klyshko, 1988, Phys.  Lett.
A, {\bf 132}, 299;
A. Zeilinger, H.J.  Bernstein,  D.M.  Greenberger,
M.A.  Horne,  and   M.
\.Zukowski, in {\it Quantum
Control and Measurement},   eds.  H.  Ezawa  and  Y. Murayama
(Elsevier, 1993); A. Zeilinger, M. \.Zukowski, M.A. Horne, H.J.
Bernstein and D.M. Greenberger, in {\it Quantum Interferometry}, eds. F.
DeMartini, A.  Zeilinger, (World  Scientific, Singapore, 1994).
\bibitem{non} A.M. Gleason, J.Math. and Mech. {\bf 6}, 885 (1957);
J.S. Bell, Rev. Mod. Phys., {\bf 38}, 447 (1966); S. Kochen and E.
Specker, J. Math. Mech., {\bf 17}, 59 (1967).
\bibitem{ZZH}
M. \.Zukowski, A. Zeilinger, M.A. Horne, Phys. Rev. A {\bf 55}, 2464
(1997).
\bibitem{MS} N.D. Mermin,
Phys. Rev. D, {\bf 22}, 356 (1980); A. Garg and N.D.
Mermin,
Phys.  Rev Lett., {\bf 49}, 901 (1982); N.D. Mermin  and G.M. Schwarz, Found.
Phys.,
{\bf 12}, 101 (1982);  M. Ardehali, Phys. Rev. D, {\bf 44}, 3336 (1991); G.S.
Agarwal, Phys.
Rev. A, {\bf 47}, 4608 (1993); K. W\'odkiewicz,
Acta Phys. Pol. {\bf 86}, 223 (1994);
K. W\'odkiewicz,  Phys. Rev {\bf A 51}, 2785
 (1995).
 \bibitem{HZ} M.A. Horne and A. Zeilinger, in {\em Symposium on  Foundations
of Modern Physics}, eds. P. Lahti and P. Mittelstaedt (World Sc.,
Singapore, 1985);  M.
\.Zukowski
and J. Pykacz, Phys. Lett.  A, {\bf 127}, 1 (1988);
M.A. Horne, A. Shimony and A. Zeilinger, Phys. Rev.
Lett.,  {\bf 62}, 2209 (1989);
 J.G. Rarity and P.R. Tapster, Phys. Rev. Lett., {\bf 64}, 2495
(1990).
\bibitem{Ivanovic}I.D. Ivanovic, J. Phys. A {\bf 14}, 3241 (1981);
W.K. Wooters, Found. Phys. {\bf 16}, 391 (1986);
J. Schwinger, Proc. Nat. Acad. Sc. {\bf 46}, 570 (1960).
\bibitem{PHD} M. Reck, Dh. D. Thesis (supervisor: A. Zeilinger) (University
of Innsbruck, 1996, unpublished).
\bibitem{MMWZZ} C.  Mattle,
M.  Michler,  H.  Weinfurter,  A.  Zeilinger   and
M. \.Zukowski, Appl. Phys. B, {\bf 60}, S111 (1995).
\bibitem{RECK} M. Reck, A. Zeilinger, H.J. Bernstein and P. Bertani, Phys.
Rev.  Lett.,  {\bf 73}, 58 (1994);
 see  also  C.  Ulrich,  Diplomathesis,  Technical
University Vienna (supervisor: A. Zeilinger), 1993 (unpublished).
\bibitem{Stenholm} I. Jex, S. Stenholm and A. Zeilinger, Opt. Comm.
{\bf 117}, 95 (1995).
\bibitem{ZZHE}
M. \.Zukowski, A. Zeilinger, M.A. Horne and A.K. Ekert, Phys. Rev.
Lett. {\bf 71}, 4287 (1993); M. \.Zukowski, A. Zeilinger and H.
Weinfurter, Ann. Phys. N. Y. Acad. Sci.{\bf 755}, 91 (1995);  J.-W.
Pan, D. Bouwmeester, H. Weinfurter and A. Zeilinger, Phys. Rev. Lett.
{\bf 80}, 3891 (1998).
\bibitem{HARALD}
G. Weihs, M. Reck, H. Weinfurter, A. Zeilinger, Opt. Lett., {\bf 21},
302 (1996).
\bibitem{BELL} J.S. Bell, Physics, {\bf 1}, 195 (1965).
\bibitem{Innsbruck} D. Bouwmeester, Jian-Wei Pan, M. Daniell,
H. Weinfurter and A. Zeilinger, Phys. Rev. Lett. {\bf 82}, 1345 (1999).


\end{references}
\end{document}